\documentclass[aps,prl,floatfix,twocolumn,tightenlines]{revtex4-1}
\usepackage{amsmath,amssymb,upgreek}
\usepackage{graphicx}
\usepackage[colorlinks]{hyperref}
\usepackage{float}

\DeclareMathOperator{\sinc}{sinc}

\begin{document}

\title{Optimal Imaging of Remote Bodies using Quantum Detectors}
             
\author{L.A.~Howard$^1$, G.G.~Gillett$^1$, M.E.~Pearce$^2$, R.A.~Abrahao$^1$, T.J.~Weinhold$^1$, P.~Kok$^2$, and A.~G.~White$^1$ }
\affiliation{$^1$Centre for Engineered Quantum Systems, School of Mathematics and Physics, University of Queensland, Brisbane, Australia}
\affiliation{$^2$Department of Physics and Astronomy, University of Sheffield, Sheffield S3 7RH, United Kingdom}
     
\date{\today}
	    	     
 \begin{abstract}\noindent
 We implement a general imaging method by measuring the complex degree of coherence using linear optics and photon number resolving detectors. In the absence of collective or entanglement-assisted measurements, our method is optimal over a large range of practically relevant values of the complex degree of coherence. We measure the size and position of a small distant source of pseudo-thermal light, and show that our method outperforms the traditional imaging method by an order of magnitude in precision. Finally, we show that a lack of photon number resolution in the detectors has only a modest detrimental effect on measurement precision and simulate imaging using the new and traditional methods with an array of detectors; showing that the new method improves both image clarity and contrast.
 \end{abstract}
             
 \maketitle

 \noindent Imaging is, at its heart, a multi-parameter metrology problem, where physical characteristics of the object are encoded as parameters in the quantum state of light. Quantum metrology studies how to best measure physical quantities in the quantum regime: particularly the measurement precision of parameters that do not have an associated quantum observable---such as phase or time---and estimating  the optimal quantum measurement observable from which the parameter can be estimated \cite{Metrology_Review}. Of special interest is the estimation of multiple parameters where the optimal measurement observables for two or more parameters may not be simultaneously co-measurable---e.g., when the observables do not commute. Quantum metrology provides bounds on the achievable precisions of these parameters and determines jointly optimal measurement strategies \cite{Ragy16}.

 It is well known that there are physical limits to the precision with which an image can be formed. The Rayleigh-Abbe limit states that the size of the smallest resolvable features is determined by the ratio of the wavelength and the numerical aperture. There are ways in which this limit can be circumvented, for example using super-resolution techniques that exploit the physical structure of the object \cite{Rust_super_resolution,betzig2006imaging,fernandez2008fluorescent}, or object illumination with entangled states of light \cite{boto00,shih01,shih2007quantum_imaging,lloyd2008quantum_illumination,Vienna_2014quantum_imaging,Brisbane2017unconditional_shot-noise}. However, in many applications---for example when the object is very far away---we cannot directly interact with the object, or illuminate it with entangled light: the quantum state of the light field is all that is accessible to the observer. Given a finite size imaging system in the far field---i.e., systems with a finite effective numerical aperture---we here investigate the best way to extract the spatial characteristics of the light source.

 \begin{figure}[b]
 \vspace{-5mm}
 \includegraphics[width=1.0 \columnwidth]{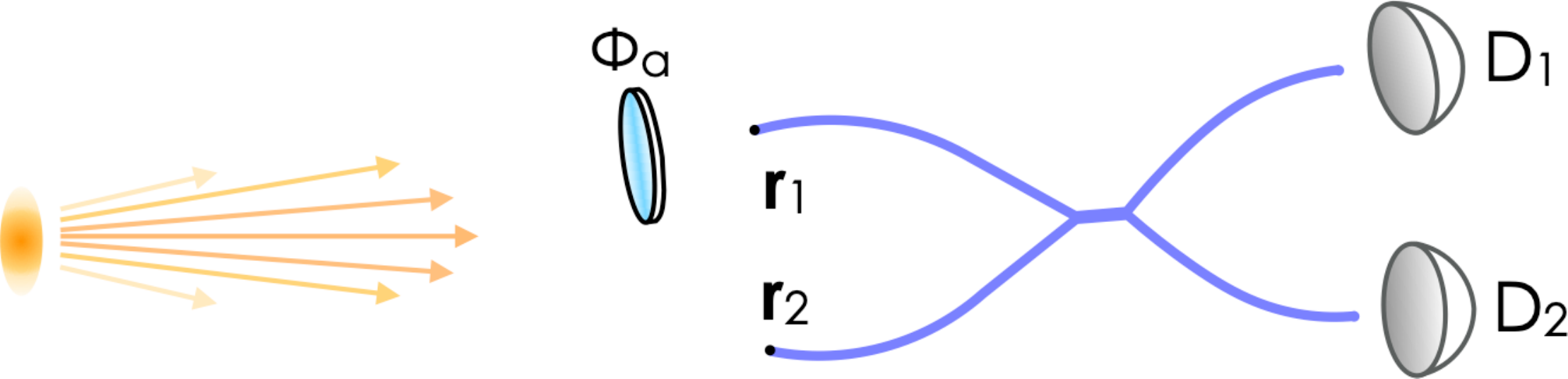}
 \vspace{-5mm}
 \caption{Schematic for the Count, Traditional, and Click schemes for estimating the complex degree of coherence (CDC) of the light field at positions $\mathbf{r}_1$ and $\mathbf{r}_2$. (This arrangement provides a general imaging procedure since the CDC is directly related to the Fourier transformation of the source distribution via the Van Cittert-Zernike theorem). The incoming light fields at \textbf{r}$_{1}$ and \textbf{r}$_{2}$ are interfered at a beam splitter, the output of which is sent to detectors $D_{1}$ and $D_{2}$. In the Count scheme the detectors are photon-number-resolving and the phase shift $\phi_a$ is random; the Traditional scheme is similar except the phase is fixed. In the Click scheme the phase is random but the detectors are not number-resolving, instead recording events if one or more photons are present. \vspace{-1mm} }
 \label{theorySchematic}
 \end{figure}

 Recently, Tsang \emph{et al}.\ showed that the far field quantum state of light retains a significant amount of information about the separation of two identical incoherent point sources, even when their angular separation approaches zero \cite{Tsang16}. Moreover, this information can be extracted with a suitable measurement \cite{Sidhu17}, for example using spatial-mode demultiplexing \cite{Tsang17}. Recently, a series of experiments demonstrated sub-Rayleigh resolution for two incoherent point sources, using image inversion interferometry \cite{Tang16,Tham17}, digital holography \cite{Paur16}, and TEM$_{01}$ heterodyne detection \cite{Yang16}. However, the retention of the spatial information seems to be restricted to highly symmetric sources \cite{Rehacek17}, and it is an open question how we can optimally extract the spatial characteristics of arbitrary sources. Possible candidates include conventional telescopes, Hanbury Brown and Twiss interferometry \cite{HBT,qunatum_img_free_electron}, or estimating higher-order correlations in the far field \cite{Oppel_super_resolution,pearce15,Genovese16}.

 In this paper, we consider the important practical case where we do not know the light source distribution, and therefore do not have a simple theoretical model whose parameters we can estimate. This requires that we measure quantities with a special relation to the source distribution, such as the complex degree of coherence (CDC). The van Cittert-Zernike theorem relates the CDC to the source distribution via a two-dimensional Fourier transform \cite{Mandel95}, which is easy to evaluate. Pearce \emph{et al}.\ showed that the CDC, $\gamma(\mathbf{r}_1,\mathbf{r}_2) {=} |\gamma| e^{i\phi}$, between two points $\mathbf{r}_1$ and $\mathbf{r}_2$ in the imaging plane can be measured nearly optimally using the setup in  Fig.~\ref{theorySchematic} \cite{Pearce2017}: The two main features are application of a varying phase, $\phi_a$, to one mode of incoming light and measurement using photon-number-resolving detectors: given the latter, we label this the ``Count'' scheme. Importantly, since the optimal method involves measurement of non-commuting observables, the Count scheme represents the optimal experimentally achievable scheme, and achieves very close to the optimal theoretical precision until $|\gamma|$ approaches approximately 0.8.

 Here, we experimentally implement the Count scheme, comparing it to two alternative schemes both of which are variations of the apparatus in Fig.~\ref{theorySchematic}. The first of these is the two-mode analog to a traditional lensing and intensity measurement setup: this ``Traditional'' scheme replaces the variable phase with a fixed phase. In contrast, the second scheme retains the variable phase but uses detectors that are not photon-number-resolving: in this ``Click'' scheme the detectors merely record an event when one or more photons are present, similar to avalanche photodiode detectors.

 Optimality of a scheme is defined via the mean squared error (MSE) in the parameters using unbiased estimators. The quantum Cram\'er-Rao bound relates the MSE matrix for these parameters to the quantum Fisher information matrix determined by the light field captured in the detectors \cite{Pearce2017,Helstrom73}. The quantum Fisher information in turn determines the optimal measurement observables, leading to the setup in Fig.~\ref{theorySchematic}. While the parameters of interest $|\gamma|$ and $\phi$ have non-commuting measurement observables, they turn out to be jointly measurable \cite{Pearce2017}.

 The coherence, CDC, is determined directly from interference fringes between two spatially separated optical modes in the far field. Light at positions $\mathbf{r}_1$ and $\mathbf{r}_2$ acquires a relative applied phase shift $\phi_a$ (using the phase shifter in Fig.~\ref{theorySchematic}), then interferes on a 50:50 beam splitter, and is finally detected by photon-number-resolving detectors $D_{1}$ and $D_{2}$, with $x$ the number of photons in detector $D_1$ and $y$ the number of photons in $D_2$. Post-selecting on different photon-number coincidence events $[x,y]$ gives rise to different interference fringes, as illustrated in Fig.~2 in the supplementary materials. (Note that these calibration fringes are not used for the experimental estimates in Figs.~\ref{phigammaNPN} and \ref{phigammaFPS}).

 The phase $\phi$ of the CDC contains the information about the position of the source relative to the optical axis connecting the source and the imaging plane. To see this, we note that a transversal shift of the source in the direction parallel to $\mathbf{r}_1 {-} \mathbf{r}_2$ produces a relative phase shift in the optical modes at $\mathbf{r}_1$ and $\mathbf{r}_2$. This results in a translation of the interference fringes. The phase $\phi$ of the CDC is equal to the applied phase $\phi_a$ at the point where the fringes are all at extremal values. For example, we can infer from Fig.~2 in the supplementary materials that $\phi$ is slightly less than $\pi$ for the calibration dataset (see the Supplementary Material for more details on the definition and measurement of $\phi$).

 The magnitude $|\gamma|$ of the CDC contains the information about the spatial extent of the source, and is equal to the visibility of the fringes,
 \begin{align}\label{gammaEqn}
  |\gamma| = \frac{I_{\rm max}^{[x,y]}-I_{\rm min}^{[x,y]}}{I_{\rm max}^{[x,y]}+I_{\rm min}^{[x,y]}}\, ,
 \end{align}
 where $I_{\rm max}^{[x,y]}$ and $I_{\rm min}^{[x,y]}$ are respectively the maximum and minimum intensity of the interference fringe for detector coincidence events $[x,y]$. To see this, we note that each single point at the source creates an interference fringe with perfect visibility and $|\gamma| {=} 1$. Incoherent extended sources at different positions then create an incoherent superposition of horizontally translated interference fringes. As the spatial extent of the source increases, the visibility---and hence $|\gamma|$---of the resulting interference fringe decreases. These relationships are formally expressed in the van Cittert-Zernike theorem (see Supplementary Material).

 To extract the values of $|\gamma|$ and $\phi$ from a measured interference fringe we use the maximum likelihood estimator (MLE), which is asymptotically efficient. This means that the variance of the MLE asymptotically approaches the Cram\'er-Rao bound for large datasets. The MLE optimises $|\gamma|$ and $\phi$ to fit the experimental data to the probability distribution $\mathrm{Pr}(x,y) {=} f_{x,y}\big( \gamma,\bar{n},\phi_a \big)$ for detecting a coincidence event $[x,y]$. This distribution is a function of $|\gamma|$, $\phi$, the average photon-number $\bar{n}$, and the applied phase $\phi_a$ and is defined explicitly in the supplementary materials.

 \begin{figure}
 \includegraphics[width=\linewidth]{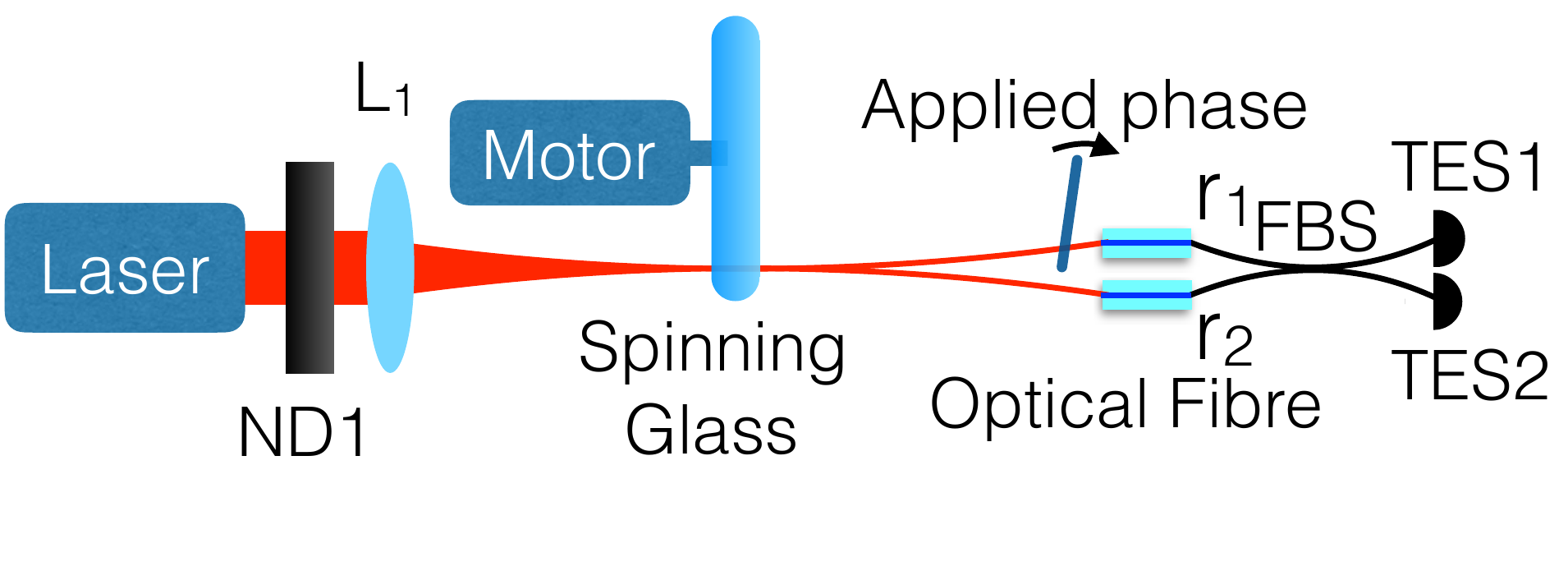}
 \vspace{-10mm}
 \caption{A 10\;kHz pulsed 820\;nm laser is attenuated with a neutral density filter (ND1) and focused on a ground glass plate; the rotation of which turns spatially coherent light into spatially incoherent (thermal) light. In the far field, the light is collected into two optical fibres $\mathbf{r_{1}}$ and $\mathbf{r_{2}}$ with a glass plate located in front of $\mathbf{r_{1}}$. As the plate rotates, it applies a relative phase shift $\phi_a$. Both arms are connected to a 50:50 fibre beam splitter (FBS). The outputs are connected to two transition edge sensors, TES1 and TES2, which measure incoming photons in the photon-number basis with single photon-number resolution and near-unit quantum efficiency. \vspace{-5mm}}
 \label{schematic}
 \end{figure}

 The experimental apparatus is shown in Fig.~\ref{schematic}. Light from a pulsed 820\;nm laser diode is spatially filtered using a single-mode fibre, and attenuated to produce a train of weak coherent states with average photon number of approximately one. The single mode is focused to a waist at a ground glass plate that rotates at approximately 5~Hz to create a spatially incoherent light source \cite{1964AmJPh..32..919M}.  The pseudo-thermal light emanating from the ground glass is collected into two single mode fibres, separated by 48~mm at $\mathbf{r}_{1}$ and $\mathbf{r}_{2}$, and located 595~mm from the ground glass plate. In front of the fibre at $\mathbf{r}_{1}$ we place a rotatable optical flat, the angle of which changes the path length of the light entering the fibre and allows for the application of the variable phase $\phi_a$. The light entering $\mathbf{r}_{1}$ and $\mathbf{r}_{2}$ is then interfered on a 50:50 fibre beam splitter. Bat-ear polarisation controllers in the fibres ensure that the polarisations in both inputs are kept equal. 

 The two outputs of the fibre beam splitter are sent to Transition Edge Sensors (TES), which are calorimeters measuring photon energy and provide true photon-number resolution when detecting monochromatic light. They also provide near-unity intrinsic detection efficiency and zero intrinsic dark counts, making them ideally suited for low-light experiments. The TES output yields a time and number resolved detection record \cite{Geoff_thesis}.

 Our setup constitutes an interferometer measuring optical coherence between $\mathbf{r}_{1}$ and $\mathbf{r}_{2}$. Stable interferometry requires that mechanical fluctuations in the position of the optical fibres be kept  within $\lambda /5\approx 40$\;nm. To achieve this, we isolate the optical fibres and optical flat within an acrylic box, with a small hole towards the laser diode to let light into the interferometer. The entire experiment is then isolated further in another box on a floating optical table. 

 Before comparing the precision of the Count scheme to the Traditional and Click schemes we discuss the accuracy of our estimates of $|\gamma|$ and $\phi$. We determine this via two methods: (i) we compare the MLE values of $|\gamma|$ and $\phi$ to the values calculated directly from the fringes; (ii) we use the MLE for $|\gamma|$ and the van Cittert-Zernike theorem to estimate the diameter of the source, and compare it to a directly measured value of the source diameter. We calculated $|\gamma| = 0.096$ using Eq.~(\ref{gammaEqn}) and we found $\phi = 4.11$\;rad from averaging the applied phase, $\phi_a$, at the extremal points for all fringes in the photon-number coincidence basis. For method (i), the visibility $|\gamma|$ was calculated from the average of only the [0,1] and [1,0] fringes. The higher order fringes were not included due to the presence of increased noise. Outliers in the data would inflate their visibility. The MLE does not suffer from this drawback. The method (i) calculated values for both $|\gamma|$ and $\phi$ are consistent with the results of the MLEs for all three schemes.
 \begin{table}[t!]
 \begin{center}
 \begin{tabular}{lcc} 
  \hline\hline
  Scheme & $|\gamma| = |\sinc (k d\, a/2D)|$  & $\mathbf{\phi} = kd\, \theta$  \\  
  \hline
  \textbf{Count} & $0.096\pm0.022$ & $4.32\pm 0.25$\\ 
  \textbf{Click} & $0.095\pm0.025$ & $4.29\pm 0.35$\\
  \textbf{Traditional}\!\!\!\!\!\! & $0.20\pm0.16$ & $4.5\pm 1.0$\\
  \hline\hline
 \end{tabular}
 \caption{Complex degree-of-coherence magnitude, $|\gamma|$, and phase, $\phi$, for the Count, Traditional and Click schemes. The size of the spot is $a$, the distance between $\mathbf{r}_{1}$ and $\mathbf{r}_{2}$ is $d$, the wave number is $k$, the distance from the source to the collection points is $D$, and the angle to the centre of the spot is $\theta$ (see Supplementary Material). Values are the average for datasets of sizes 1000 to 10\,000.} 
 \label{averageValues}
 \vspace{-9mm}
 \end{center}
 \end{table}

 For method (ii) we measured the source diameter, which is equal to the spot size of the beam incident on the ground glass plate, see Fig.~\ref{schematic}. The ground-glass plate was placed within $\pm0.25$\;mm of the beam waist. (The uncertainty in position is due to a small amount of precession of the rotating plate). Using a beam profiler we measured the spot size at the waist to be $15.3 \pm 0.1~\upmu$m and the spot size 0.25\;mm from the waist to be $18.0 \pm 0.1~\upmu$m. We therefore expect to estimate a source diameter in the range $15.3$\,-$18.0~\upmu$m. Using the van Cittert-Zernike theorem and our estimated visibility we estimate the source diameter to be $16.5 \pm 0.5~\upmu$m, agreeing well with the predicted range of diameters (see the Supplementary Material for details).

 Table~\ref{averageValues} summarises our estimates of the complex degree-of-coherence magnitude, $|\gamma|$, and phase, $\phi$, using the Count, Traditional, and Click schemes, averaged over dataset sizes from 100 to 10000 points. For $|\gamma|$, the Count scheme is respectively 7.3 and 1.1 times more precise than the the Traditional and Click schemes; for $\phi$, the Count scheme is respectively 4.0 and 1.4 times more precise than the the Traditional and Click schemes. Moving beyond these averages, for the 10,000 point dataset, the Count scheme is over an order-of-magnitude more precise for $|\gamma|$, and four times more precise for $\phi$, than the Traditional scheme. These results also apply to comparisons between the Click and Traditional schemes since Fig.~\ref{phigammaNPN} displays that the Count and Click schemes are of approximately equivalent precision for larger dataset sizes. These results demonstrate that the Count and Click schemes are significantly better than the Traditional scheme, corroborating the result by Pearce \emph{et al}.\ that the Count scheme is near-optimal among non-collective measurements \cite{Pearce2017}. Estimates for $|\gamma|$ and $\phi$ based on various data set sizes are shown in Figs.~\ref{phigammaNPN} and \ref{phigammaFPS}. They reveal that the Count scheme converges more quickly around the true values of $|\gamma|$ and $\phi$ than the Traditional scheme.
 \begin{figure}[t]
 \includegraphics[width=8.5cm]{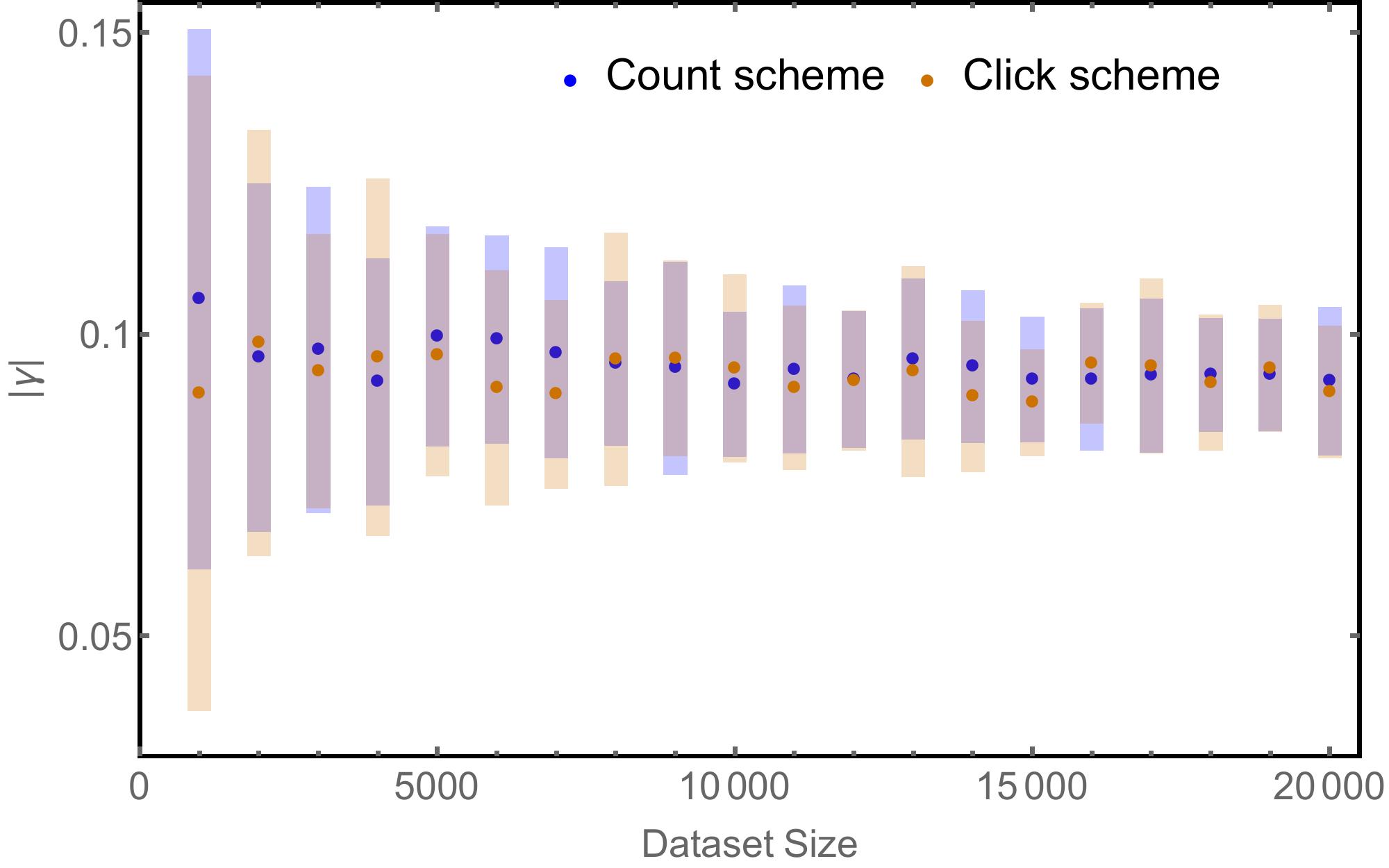} \\
 \vspace{1mm}
 \hspace{3mm}\includegraphics[width=8.1cm]{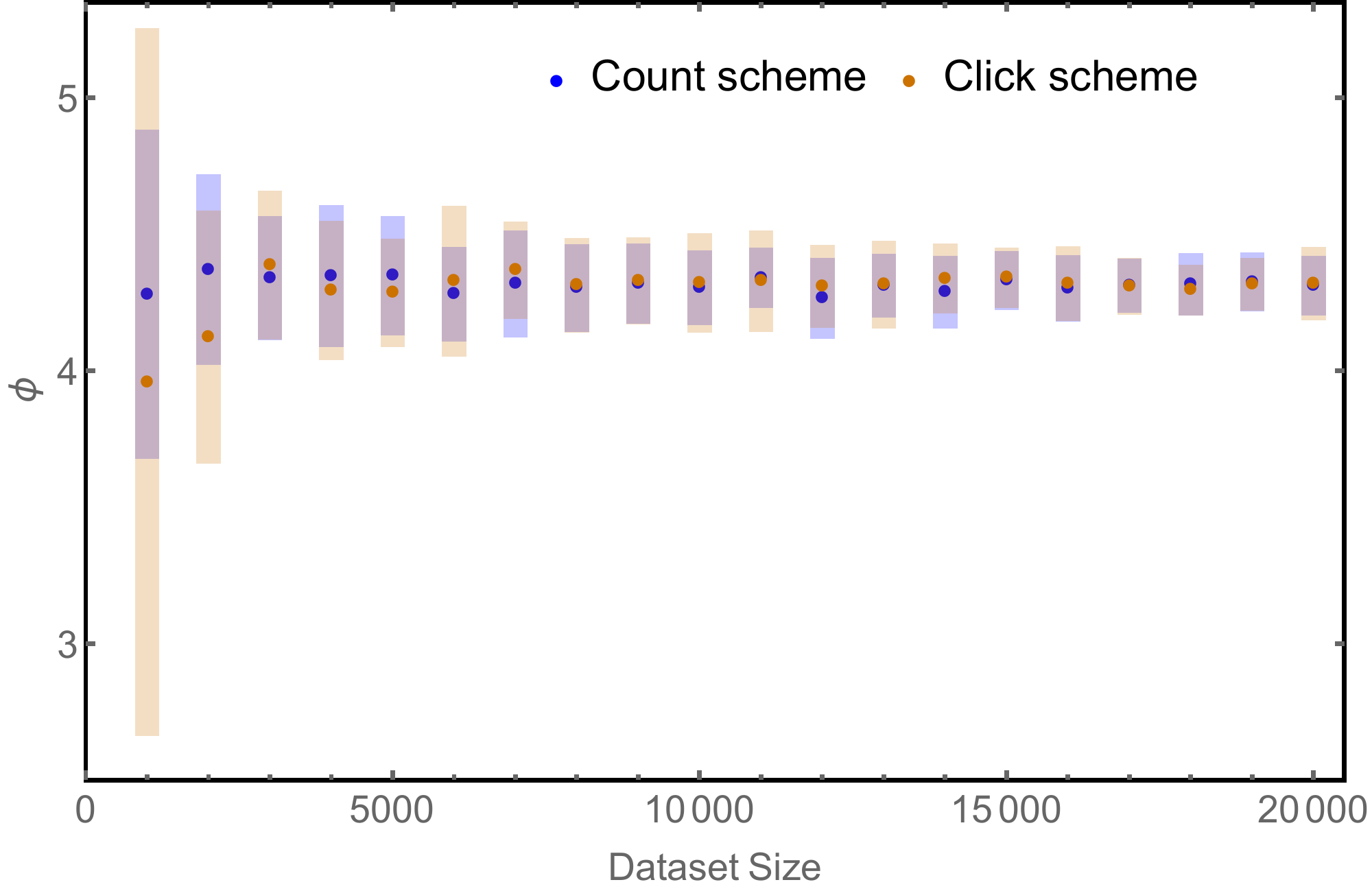}
 \vspace{-2mm}
 \caption{\emph{Count vs Click}. Complex degree-of-coherence versus dataset size: \emph{Top} magnitude, $|\gamma|$. \emph{Bottom} phase, $\phi$. Each point is the mean of 50 trials with the shading representing the standard deviation of the 50 trials. Blue dots and shading are for the Count scheme  and orange dots and shading are for the Click scheme. For small data sets, the Count scheme has a clear advantage over the Click scheme; both perform well at large data set sizes.\vspace{-5mm}}
 \label{phigammaNPN}
 \end{figure}

 \begin{figure}[t]
 \vspace{-3mm}
 \includegraphics[width=8.3cm]{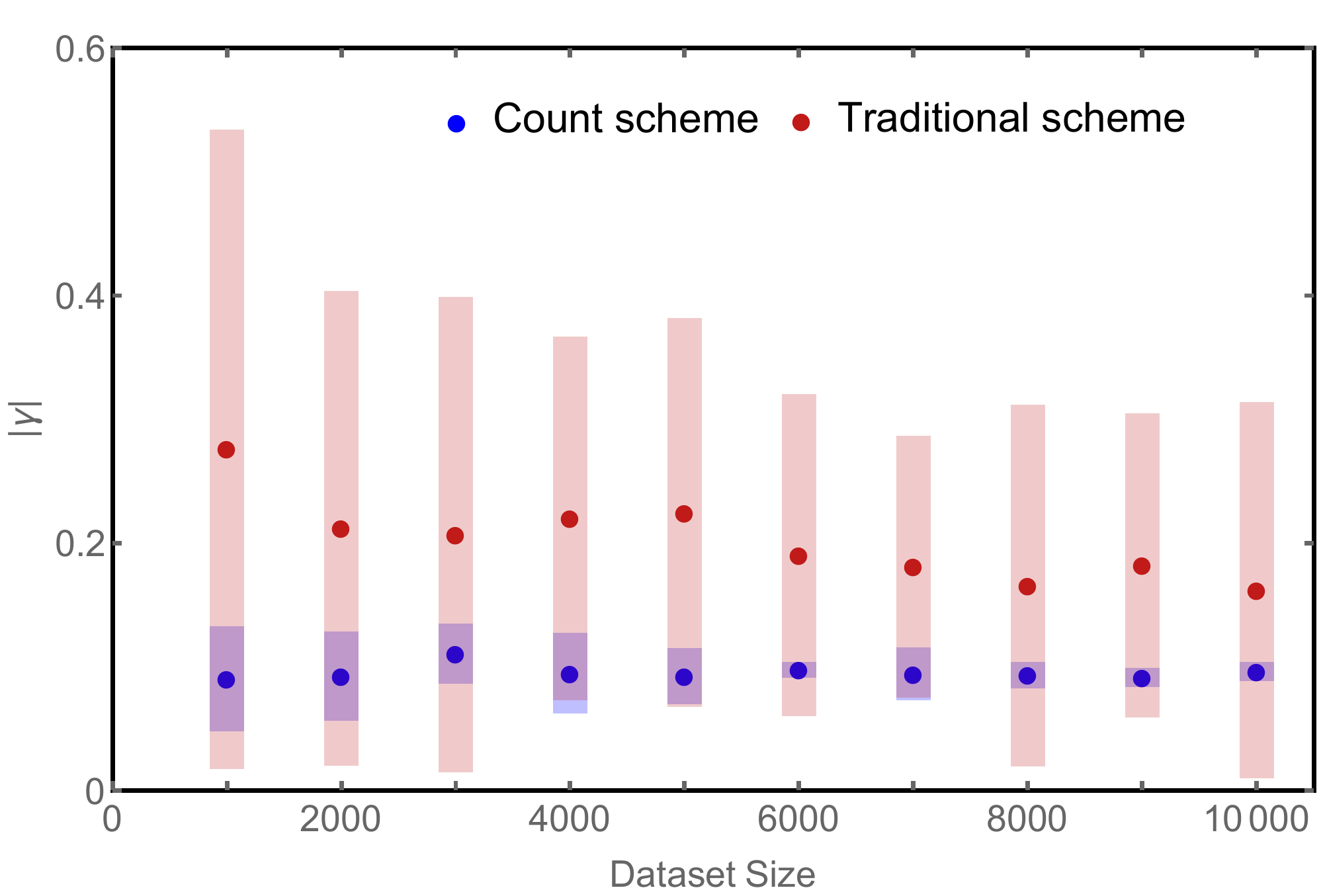} \\
 \vspace{-2mm}
 \hspace{2mm}\includegraphics[width=8.1cm]{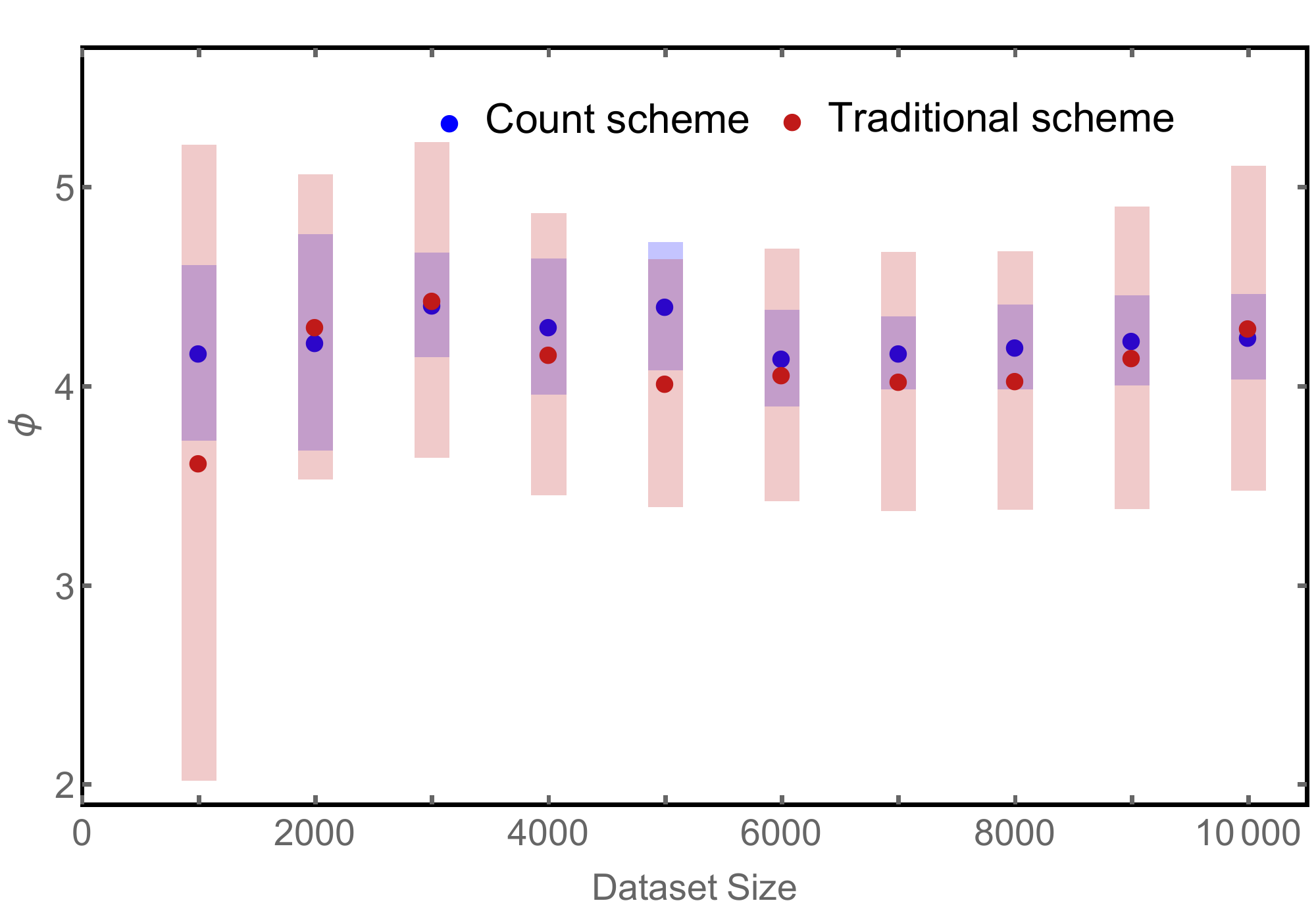}
 \vspace{-2mm}
 \caption{\emph{Count vs Traditonal}. Complex degree-of-coherence versus dataset size: \emph{Top} magnitude, $|\gamma|$. \emph{Bottom} phase, $\phi$. Each point is the mean of 20 trials with the shading representing the standard deviation of the 20 trials. Blue dots and shading are for the Count scheme and red dots and shading are for the Traditional scheme. The Count scheme clearly outperforms the Traditional scheme for all dataset sizes. \vspace{-2mm}}
 \label{phigammaFPS}
 \end{figure}

In the Traditional scheme we see a consistent bias in the $|\gamma|$ value estimates, relative to the Count or Click scheme. This is due to the uncertainty in $|\gamma|$ for the Traditional scheme being greater than the difference between the true $|\gamma|$, $0.096$, and the lower limit of $|\gamma|$, zero. This results in truncation of some MLE estimates smaller than $|\gamma|$, causing inflation of the mean $|\gamma|$ estimation. 

 Practical application of the Count scheme to the imaging of objects with arbitrary spatial configurations will require a 2-dimensional array of photon-number-resolving detectors and phase varying elements. This setup could be realised by pairing a recent implementation of an array of photon-number-resolving detectors \cite{Number_resolving_Optica_2017} along with a liquid crystal spatial light modulator to dynamically vary the phase of the light entering each detector. A reconstructed image is then formed by measuring the CDC between all detector pairs in the array and calculating the source distribution via the van Cittert-Zernike theorem. 
 To demonstrate this method we simulated imaging of a complex object with a 26$\times$26 array of detectors using the Count and Traditional methods (see Fig.~\ref{imaging}). The simulation shows that the Count scheme results in images with improved contrast and clarity over traditional imaging methods.
 Results from the Click scheme (see Fig.~\ref{phigammaNPN}) shows that for large data sets, substituting the array of photon-number-resolving detectors for an array of non-photon-number-resolving detectors does not result in a large decrease in precision, and may be a more economical alternative. 
 
\setlength{\fboxsep}{0pt}
\setlength{\fboxrule}{2pt}

\begin{figure}[t!]
 \begin{center}
 \begin{tabular}{cc} 
 
 \fbox{\includegraphics[scale=0.4]{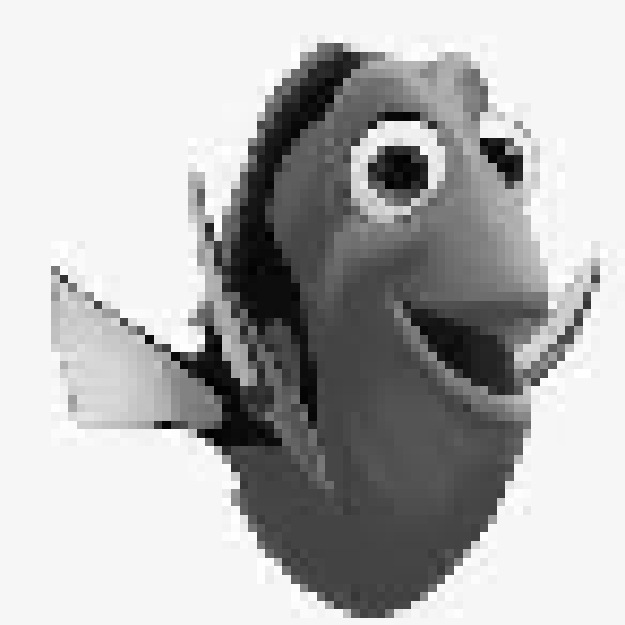}}  & 
 \hspace{-8.0mm} \vspace{0.05cm} \fbox{\includegraphics[scale=0.4]{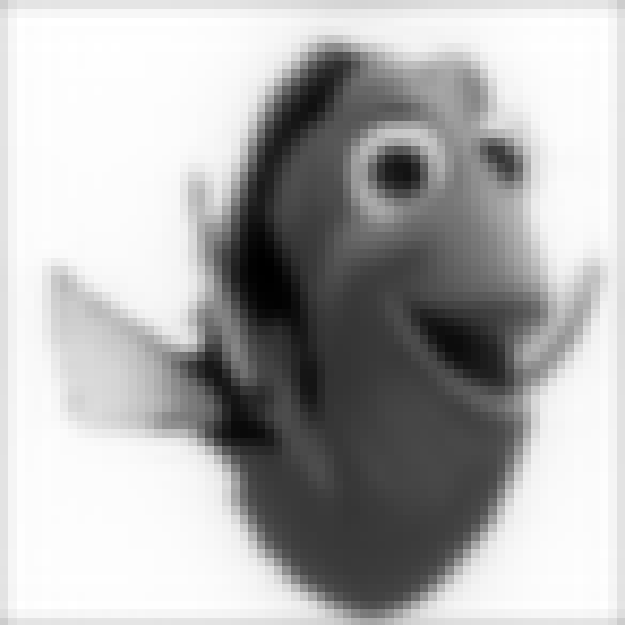}}  \\  
 
  \fbox{\includegraphics[scale=0.4]{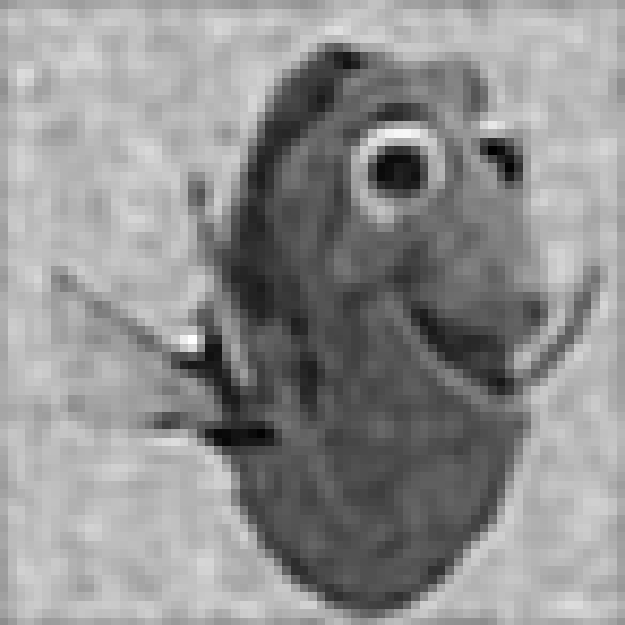}} & \hspace{-8.0mm} \fbox{\includegraphics[scale=0.306]{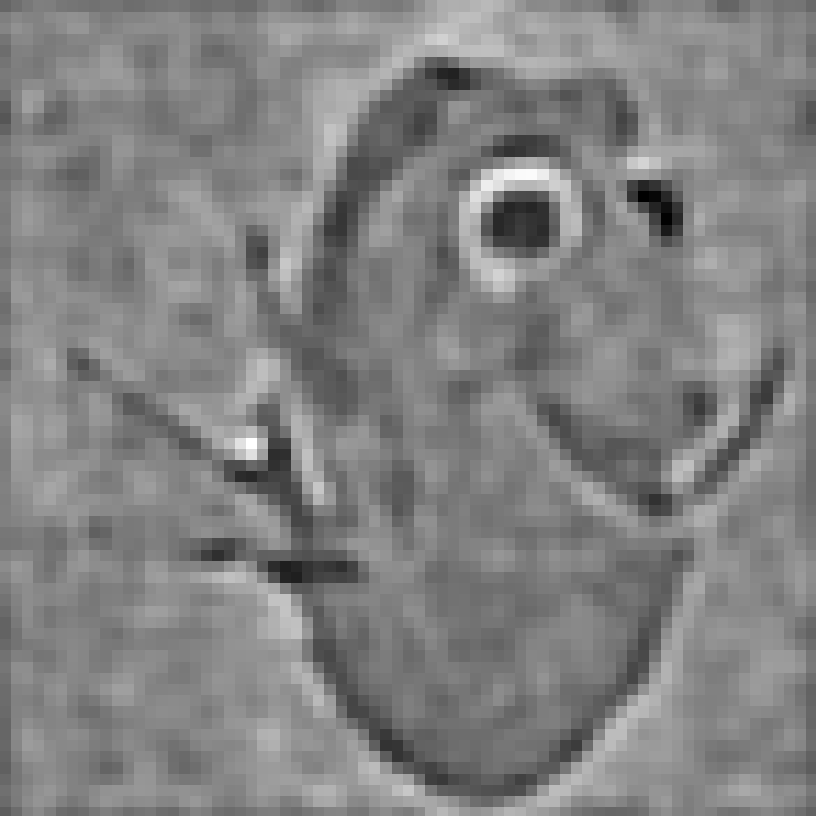}}  \\ 
  
 \end{tabular}
 \caption{Simulated comparison of images reconstructed using  the Count and Traditional schemes for a 26$\times$26 array of detectors. The Click scheme is not shown due to it, visually, looking very similar to the Count scheme {Top left}: the original image before reconstructive imaging. {Top right}: reconstructed image in a noiseless regime, revealing the theoretical limits of the method.
 {Bottom left}: reconstructed image based on our Count scheme.
 {Bottom right}: reconstructed image based on the Traditional scheme. A detailed description of the simulation is given in the Supplementary Material.} 
 \label{imaging}
 \vspace{-9mm}
 \end{center}
 \end{figure}
 
 The Count and Click schemes, in contrast to intensity methods such as Hanbury Brown and Twiss interferometry, allows for estimation of both $|\gamma|$ and $\phi$, and hence total reconstruction of the source image (see Fig.~\ref{imaging}). This, combined with their optimality in precision, their relative simplicity, and the ubiquity of interferometers in modern sensing and imaging technology, means the Count and Click schemes have many potential applications. For example in astronomy these schemes open avenues to improved imaging of stellar bodies, with the Count scheme having a particular advantage due to its precision at low dataset sizes. Extension of the scheme to multiport interferometry \cite{Resch2007} and the associated phase super-resolution may make possible imaging of previously inaccessible smaller bodies such as exoplanets, moons and asteroids. In biology and medical imaging---through incorporation into interferometric reflectance schemes that detect reflected thermal light shone onto the source \cite{Biological}---the Count and Click schemes can provide optimal imaging of small biological entities. 

 \paragraph{Acknowledgments} \--- The authors would like to acknowledge helpful discussions from Marcelo Almeida. This work has been supported by: the Australian Research Council (ARC) Center of Excellence for Engineered Quantum Systems (EQUS, CE170100009); the Engineering and Physical Sciences Research Council (EPSRC) via grant EP/N014995/1 and the Quantum Communications Hub (EP/M013472/1); and the University of Queensland by a Vice-Chancellor's Senior Research and Teaching Fellowship for AGW.

 \bibliographystyle{ieeetr}
 \bibliography{bib}
 
 \newpage
 
 \onecolumngrid
 
 \section*{Supplementary Material}

\begin{center}             
\begin{figure}[h]
\hfil\includegraphics[width=8.39cm]{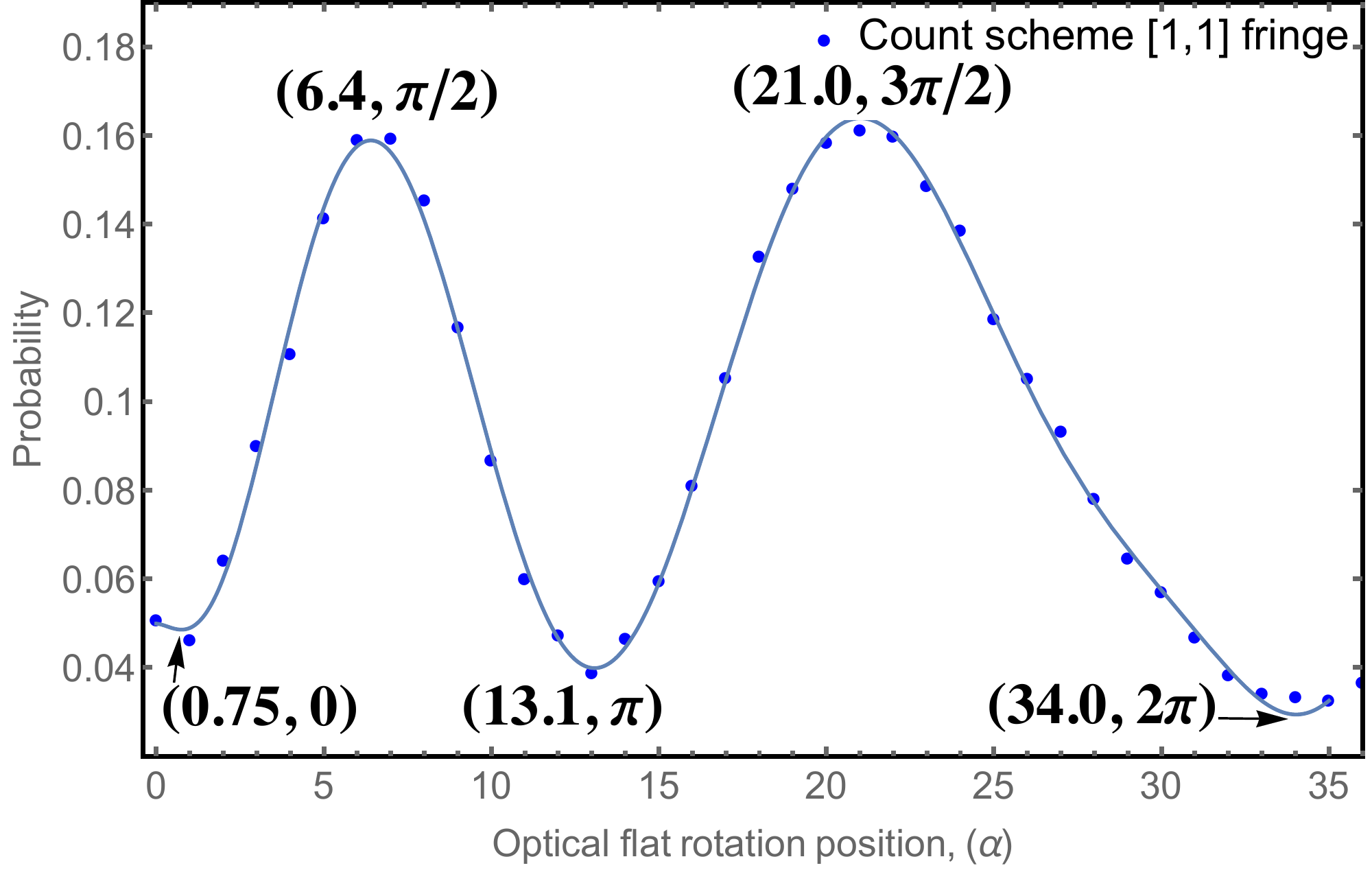} 
\includegraphics[width=8.3cm]{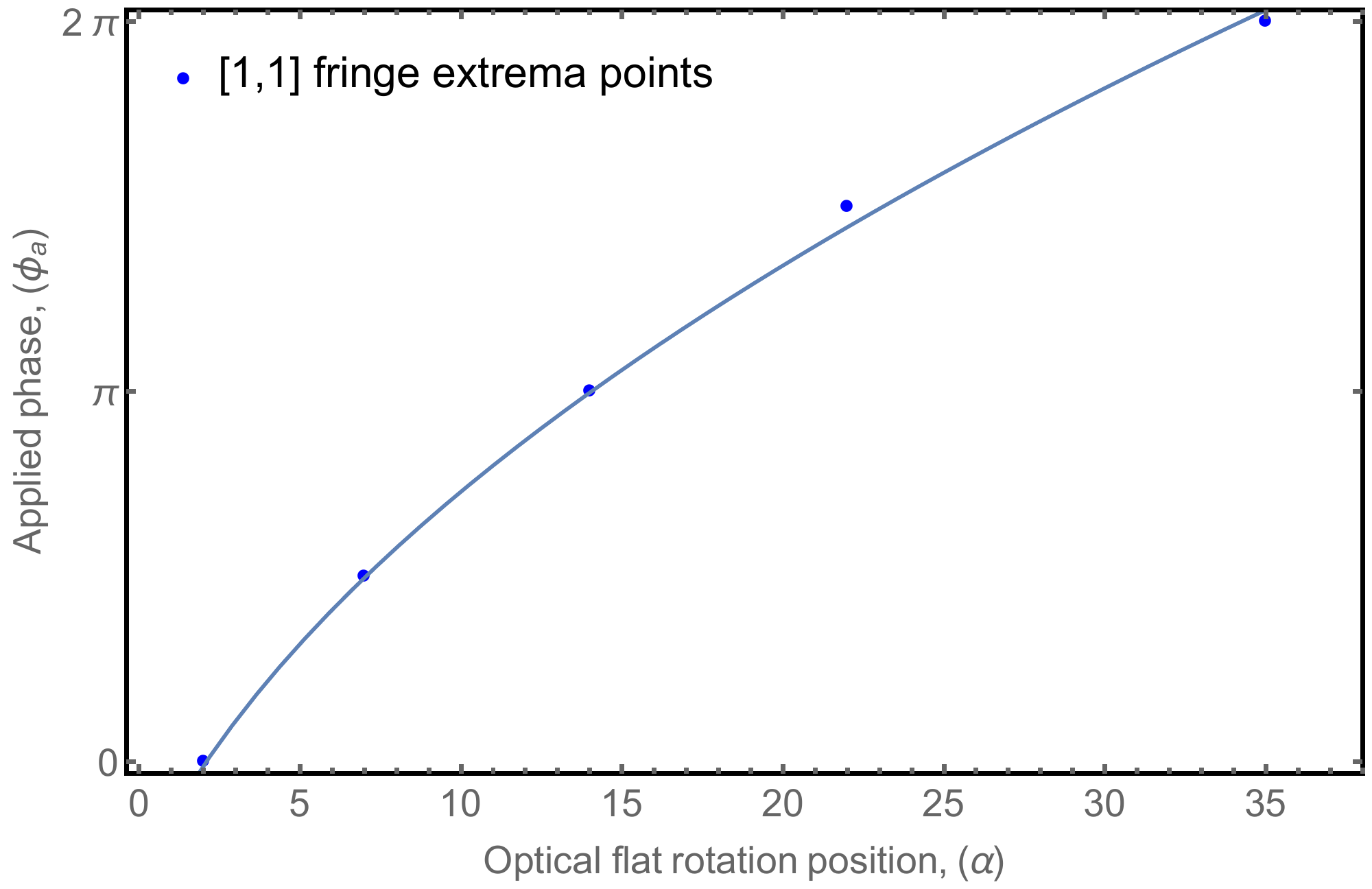}
\caption{\textbf{Left}: A [1,1] interference fringe. The horizontal axis displays different rotations of the optical flat, which acts to apply a varying phase to the light entering $\textbf{r}_{1}$. The various rotation positions are unitless. The optical flat rotational positions corresponding to extrema and the applied phase at that extrema are also labelled. \textbf{Right}: A curve fitted to the labelled extrema points points shown in the left plot provides a relationship between optical flat rotation position and applied phase. Using this relationship the phase entering $\textbf{r}_{1}$ can be determined at any rotational position of the optical flat. The equation of the fitted curve is $\phi_{a}=-2.6+1.5\sqrt{1.0+\alpha}$. }
\label{phaseCharcterisation}
\vspace{-0.5cm}
\end{figure}
\end{center}

\noindent
\textbf{A. Phase characterisation}. The rotation of the optical flat in front of $\mathbf{r}_1$ was used to set the phase shift $\phi_a$. The phase versus angle calibration for the optical flat was obtained by inputting a coherent state into the light collectors to create bright fringes, which were then measured with the photon-number-resolving detectors. We first examine the [1,1] fringe, due to its relatively high counts---i.e., good signal to noise ratio---and because it exhibits twice the frequency of the fringes that do not represent coincidence events---i.e., a combination of the Hong-Ou-Mandel effect and the increased sensitivity of a two-photon NOON state. We arbitrarily assign to the rotation position of the first [1,1] extremum an applied phase of zero. The rotation position of the second [1,1] extremum will occur when the applied phase has increased to $\pi/2$. By the third extremum the applied phase has increased to $\pi$, and so on. These extrumums are defined as the minimum and maximum points of a polynomial fitted to the fringe; the fringe, corresponding fitted polynomial and extrumum points are shown in the left plot of Fig.~\ref{phaseCharcterisation}. A final function giving the relationship between the optical flat rotation position, $\alpha$ and applied phase of light entering $r_{1}$, $\phi_{a}$, is given by fitting a curve to the extrumum points; this curve and extrumum points are plotted on the right side of Fig.~\ref{phaseCharcterisation}. 

After performing the phase characterisation, a thermal state is inserted in the setup and the Count scheme is used to estimate the CDC of the source. During the application of the Count scheme it is assumed that the applied phase is changing with optical flat rotation in the same manner as during the phase characterisation process. After application of the Count scheme we reinsert a coherent state and re-perform the phase characterisation in order to identify if there has been substantial drift of optical apparatus during application of the Count scheme. This drift may occur due to pressure gradients around the experimental setup. Based of the amount of drift present and the duration of the measurement, a decision is then made whether to keep the dataset. Typically, a drift of more than 10 degrees per hour resulted in discarding the dataset.

\vspace{-0.5cm}

\noindent
\textbf{B. Data Analysis}. Measurements of coincidences in the number basis were taken using 35 different phases between zero and $2\pi$ radians. An MLE algorithm was then used to estimate $|\gamma|$ and $\phi$ by fitting a probability function (see section E) to the measured coincidences as the phase was varied. The sample of coincidences used in the MLE process was varied in size from 1000 to 10,000 points, when comparing the Traditional and Count schemes and from 1000 to 20,000 points when comparing the Count and Click schemes. The discrepancy in the range of datasets between the two comparisons is due to the Traditional scheme measurements being a subset of the Count scheme data at one particular phase and hence containing less measurement data. A number of trials were taken at each dataset size, each using different samples from the total measurement dataset. Before any sample is taken from the total measurement set, the set is randomised and samples for each trial are selected such that no data point is used in more than one sample. The randomisation ensures each sample contains a mixture of measurements with respect to the varied phase, and time over which the measurements were taken. Due to the Traditional dataset being a subset of the Count dataset at one phase, the time over which its measurements were taken was approximately 1/35 the time over which the Count and Click schemes were taken. For this reason we can assume that the Traditional dataset is less affected by optical drift. Despite this, it performs much worse than either the Click or Count schemes. All three methods, Count, Traditional and Click use the same one dataset, with the Traditional using a subset with one phase, Click ignoring photon-number information and Count using all phases and photon-number information. There were no outliers in the dataset and as such no points were removed at any time from the calculations.\\

\begin{figure}
 \centerline{
 \vspace{-4mm}
 \hspace{0mm}
 \includegraphics[width=0.31 \columnwidth]{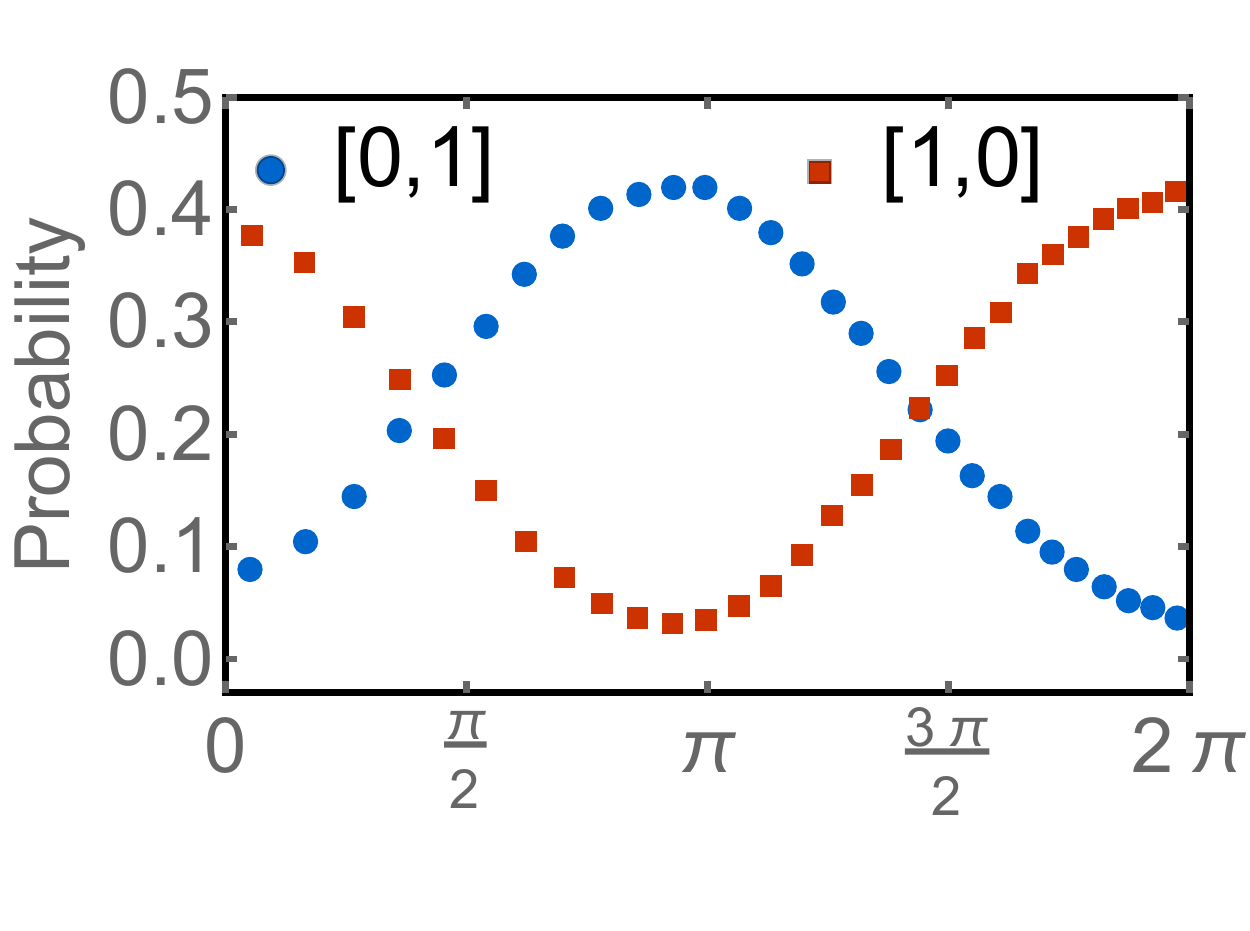}
 \includegraphics[width=0.3 \columnwidth]{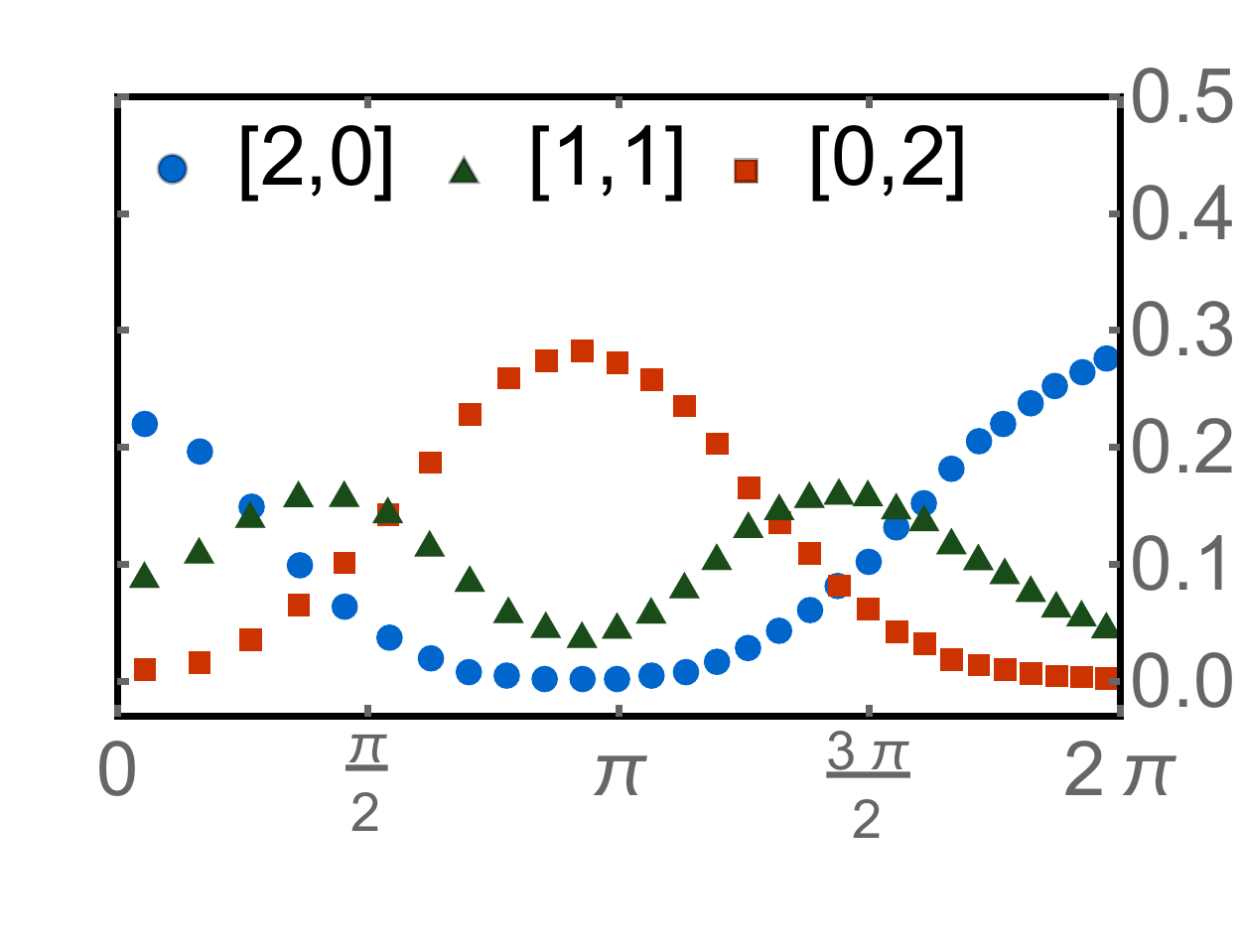}
 \hspace{0mm}
 }
 \vspace{-2mm}
 \centerline{
 \hspace{-1mm}    
 \includegraphics[width=0.31 \columnwidth]{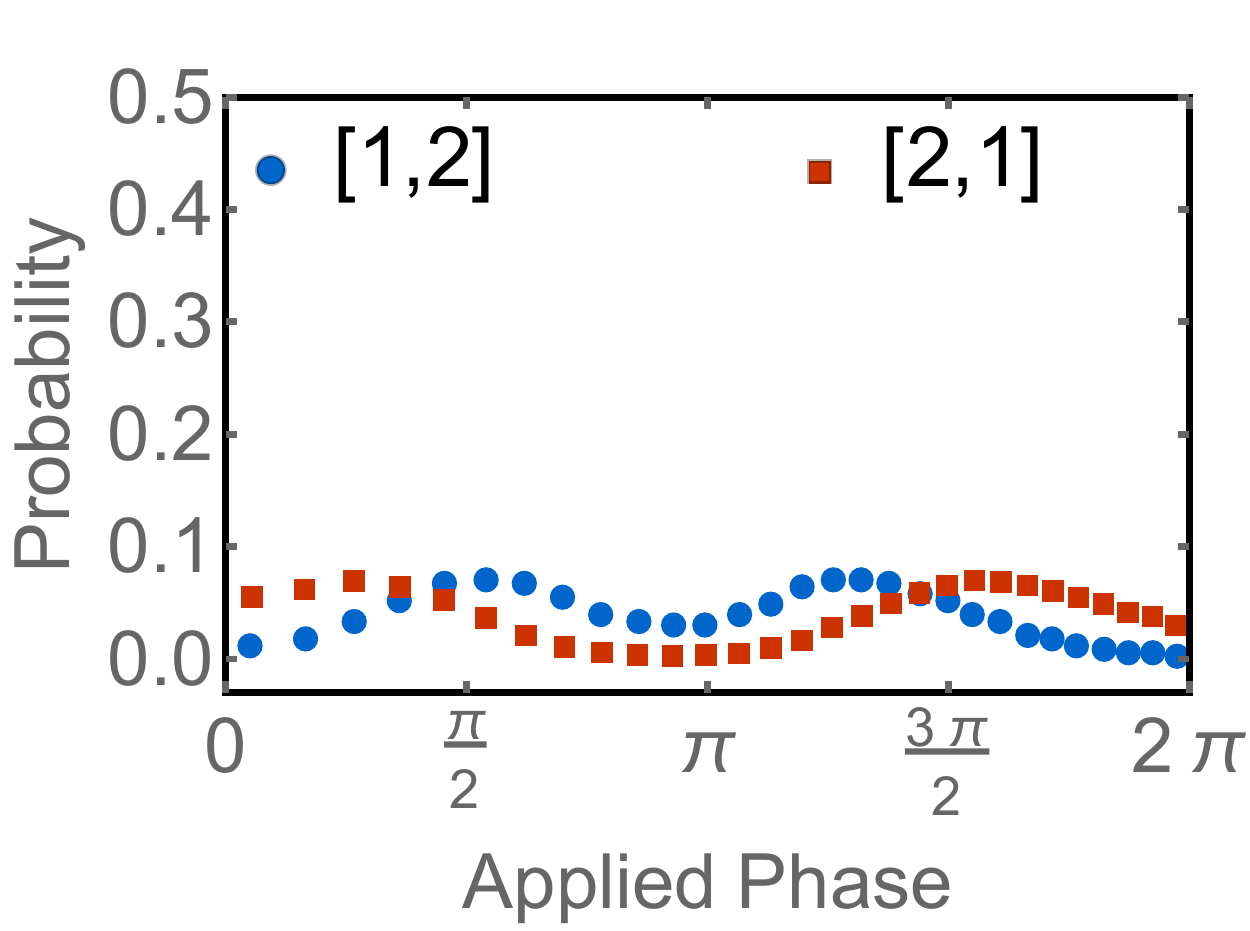}
 \includegraphics[width=0.3 \columnwidth]{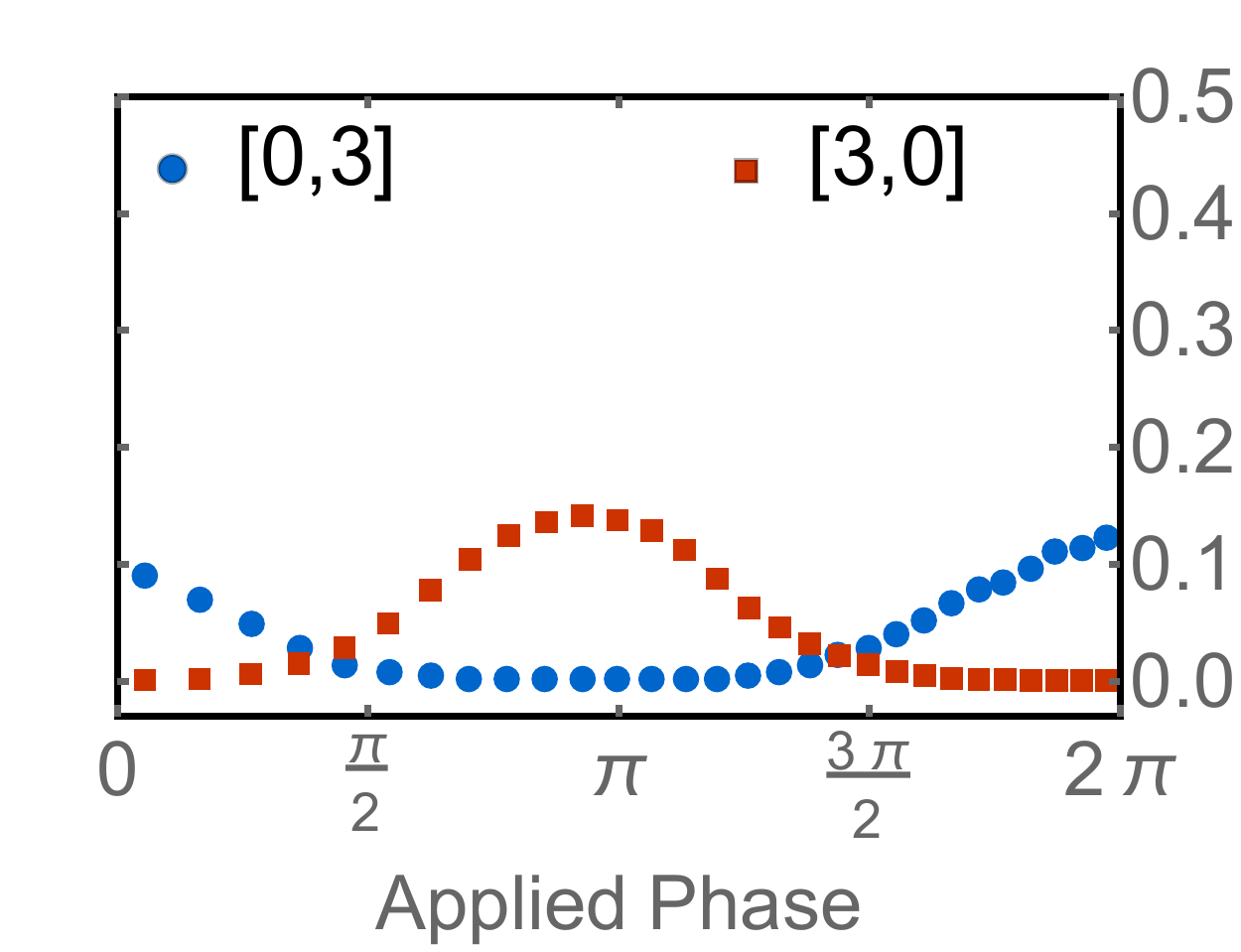}
 }
 \vspace{-2mm}
 \caption{\emph{Calibration interference fringes}. Light collectors at $\mathbf{r}_1$ and $\mathbf{r}_2$ imaged a weak, coherent-state, light source. The plots show various photon-number coincidence events [$x$,$y$]---$x$ photons in detector $D_1$, $y$ photons in detector $D_2$---versus  the applied phase $\phi_a$. Significant visibility, and therefore information, remains as the total detected photon-number, $x{+}y$, increases.
 \vspace{-6mm}}
 \label{photonFringes}
 \end{figure}

\noindent
\textbf{C. Source size and angle estimation}. Our determination of the beam diameter from the visibility calculations used the expression 
$|\gamma|(d) {=} \exp[-d^{2}/(2\sigma_{d}^{2})]$, Where $\lambda$ is the wavelength, $|\gamma|$ is the visibility of the fringes, $L$ is the distance from the source to the detectors, $d$ is the distance between the detectors, $\sigma_{d} {=} \lambda L/(2\pi \sigma_{y})$ is the spatial frequency characteristic, and $\sigma_{y}$ is the standard deviation of a Gaussian source.
\begin{figure}[h]
\includegraphics[width=8.5cm]{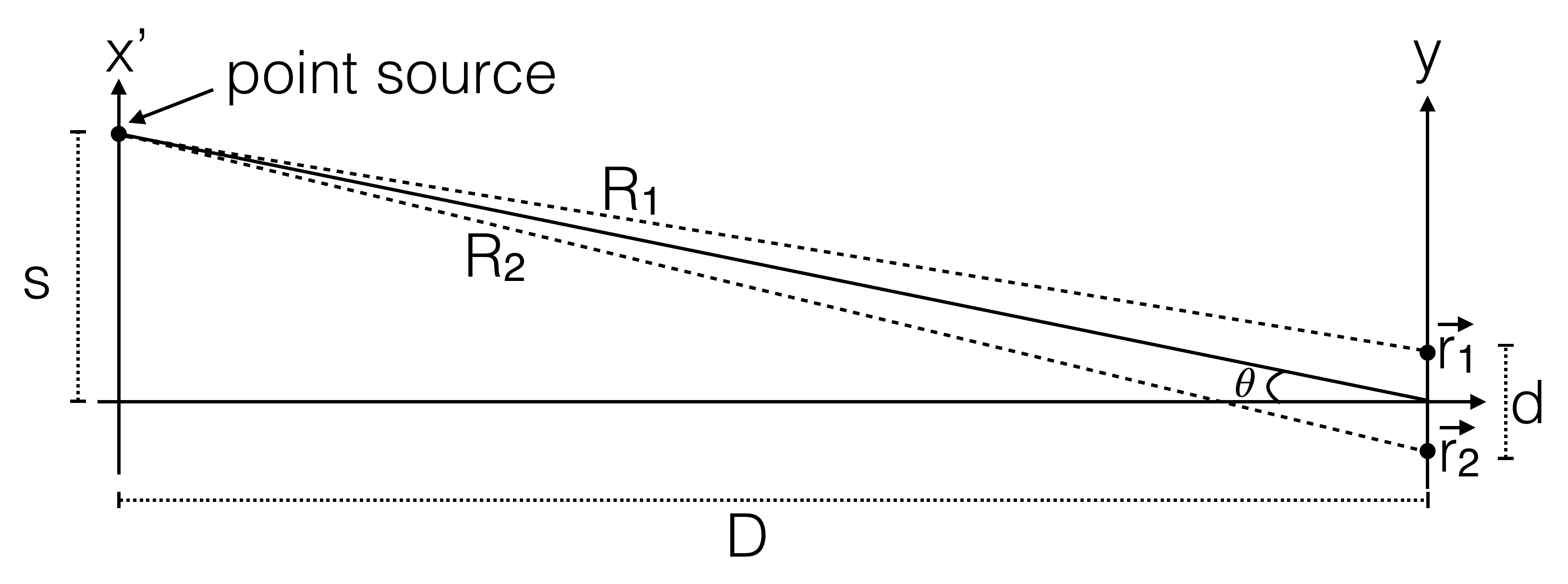}
\caption{Geometry of the source distribution and the photodetectors. Two detectors at $\mathbf{r}_1$ and $\mathbf{r}_2$ are placed along the $y$-axis perpendicular to the optical axis. The incoherent point source makes an angle $\theta$ with the optical axis at a distance $D$ from the detector plane. We seek the relationship between the phase $\phi$ of the CDC and the angle $\theta$.}
\label{fig:CDCphase}
\end{figure}

To define the relationship between the estimated parameter $\phi$ and the angle to the source we first consider an incoherent point source distribution in the object plane in one dimension, given by the intensity distribution $I_{s} {=} I_0 \delta(x' {-} s)$, with $I_0$ the total intensity of the source and $\delta(x' {-} s)$ the Dirac delta function. Up to a constant factor, the complex degree of coherence is then given by [25],
\begin{align}
\gamma \propto \int_S I_{s}(\vec{r'})e^{ik(R_{2}-R_{1})} d\vec{r'}\, ,
\end{align}
with $k$ the wave number of the light emitted by the point source. The distances $R_1$ and $R_2$ are shown in Fig.~\ref{fig:CDCphase}, and are given by,
\begin{align}
 R_1^2=D^2+\left(s-\frac{d}{2}\right)^2 \quad\text{and}\quad R_2^2=D^2+\left(s+\frac{d}{2}\right)^2\, .
\end{align}
Using the approximation $\sqrt{1+x} \approx 1+\frac12 x$ for $x \ll 1$ and $\frac{s\pm\frac{d}{2}}{D} \ll 1$, we find that for a point source at an arbitrary position $x'$ along the vertical source axis,
\begin{align} 
 R_2 - R_1 = \frac{x'd}{D}\, .
\end{align}
This allows us to calculate the phase of the CDC from,
\begin{align}
 \gamma \propto \int_{-\infty}^{\infty} \delta(x'-s) e^{{ikx'd}/{D}} dx' = e^{\frac{iksd}{D}} = e^{i\phi}\, ,
\end{align}
with $\phi$ the phase of $\gamma = |\gamma| e^{i\phi}$. This immediately yields $\phi={ksd}/{D}$. The physical angle $\theta$ is given by,
\begin{align}
 \theta = \arctan\left(\frac{s}{D}\right) \approx \frac{s}{D},
\end{align}
which leads to the relationship $\phi=kd\, \theta$. The phase of the CDC is therefore directly proportional to the angle of the source position, modified by a factor $kd$ that depends on the wavelength $\lambda = 2\pi/k$ of the light and the distance $d$ between the detectors. As expected, shorter wavelengths and larger detector separation will lead to an increased sensitivity in the position angle $\theta$ when measuring the phase of the CDC.

For a uniform source extending from $x' = s-a/2$ to $x' = s+a/2$ with integrated intensity $I_0$, the complex degree of coherence is proportional to,
\begin{align}
 \gamma(d) \propto \int_{s-\frac{a}{2}}^{s+\frac{a}{2}} e^{ikx' d/D} dx' = \sinc \left( \frac{kd\, a}{2D} \right)e^{iksd/D}\, . 
\end{align}
As the size of the source $a$ grows, the magnitude $|\gamma|$ is reduced, and $|\gamma|$ therefore gives a direct measure for $a$, given knowledge of $I_0$, $k$, $d$, and $D$. Next, we show that this information can be extracted from the visibility of the fringes.

Consider the intensity $I$ in the far field due to two points, 1 and 2, in the source plane. According to standard coherence theory (see Section 4.3.1 in Mandel and Wolf [25]) this is given by,
\begin{align}
 I = I_1 + I_2 + 2\sqrt{I_1 I_2} |\gamma| \cos\phi\, ,
\end{align}
where $I_1$ and $I_2$ are the average intensities in the far field due to the two points, respectively. The visibility of the fringes in the far field can then be calculated as,
\begin{align}
 \mathcal{V} = \frac{I_{\rm max} - I_{\rm min}}{I_{\rm max} + I_{\rm min}} = \frac{2\sqrt{I_1 I_2}}{I_1+I_2}\, |\gamma|\, ,
\end{align}
where $I_{\rm max}$ and $I_{\rm min}$ are the maximum and minimum intensities in the far field corresponding to $\cos\phi {=} +1$ and $\cos\phi {=} -1$, respectively. Assuming a uniform intensity distribution $I_1 {=} I_2$, we can immediately identify the visibility of the fringes with the magnitude of the CDC.\\

\noindent
\textbf{D. Reduced Chi-squared analysis}. In order to investigate how well the estimated values of $\phi$ and $|\gamma|$ fit our observed data, we performed reduced $\chi ^{2}$ tests between our expected and measured fringes. Our expected fringes were calculated using estimated values of $\phi$ and $|\gamma|$ from the thermal dataset used in figures 4 and 5. Specifically we calculated $\chi ^{2}$ for the  [0,1], [1,0], [1,1], [0,2], and [2,0] interference fringes, which accounted for 97\% of the data in the dataset. There were 32 degrees of freedom used for each calculation. The investigation found that overall the measured fringes matched the theoretical expectations well. 

\begin{center}
\begin{table}[h!]
\begin{tabular}{lc} 
 \hline\hline
 Fringe & Reduced $\chi^{2}$\\  
 \hline
 \textbf{[0,1]} & $0.7$\\ 
 \textbf{[1,0]} & $0.7$\\
 \textbf{[1,1]} & $1.2$\\
 \textbf{[0,2]} & $1.1$\\
 \textbf{[2,0]} & $1.2$\\
 \hline\hline
\end{tabular}
\caption{Reduced $\chi^{2}$ values for datasets of 2000 points for the [0,1], [1,0], [1,1], [0,2] and [2,0] fringes.}
\label{chiValues}
\vspace{-0.0cm}
\end{table}
\end{center}

\noindent
\textbf{E. Formula for the maximum likelihood estimator (MLE)}. The probability function $P_{r}(x,y)$ used to calculate the maximum likelihood estimator (MLE) is given by [23],
\begin{align}
 P_{r}(x,y)  = \sum_{n_{1}=0}^{x+y} \frac{p_{in}(n_{1},x+y-n_{1})}{n_{1}!(x+y-n_{1})!} \frac{x!y!}{4^{x+y}} \left | \sum_{j=0}^{x} (-1)^{j} \binom{n_{1}}{j} \binom{x+y-n_{1}}{x-j}(1-e^{-i\phi})^{x+n_{1}-2j}(1+e^{-i\phi})^{x-n_{1}+2j} \right |^2,
\end{align}
where
\begin{equation}
p_{in}(n_{1},n_{2})=\frac{z_{1}^{n_{1}}}{(1+z_{1})^{n_{1}+1}} \frac{z_{2}^{n_{2}}}{(1+z_{2})^{n_{2}+1}}\, ,
\end{equation}
and $z_{1}= \bar{n} (1-|\gamma|)$, $z_{2}= \bar{n} (1+|\gamma|)$ and $x$ and $y$ are the number of photons in detectors $D_{1}$ and $D_{2}$ respectively.

\bigskip

\noindent
\textbf{F. Imaging simulation}. In order to demonstrate the effect of noise in the imaging process, we used a  picture of Dory in grey (see Fig.~6 of the manuscript, top left), and assumed a $26\times26$ detector array to reconstruct the image. We used the van Cittert-Zernike theorem to calculate the CDC for each detector pair. We chose a wavelength $\lambda = 700$~nm, a separation between pixels of $0.7~\upmu$m, and a distance between the source and the detector plane of $8.67$~m. We then added random Gaussian noise to the calculated CDC for each detector pair, the magnitude of which was chosen according to the experimental values of the uncertainty in the CDC (Table~1 of the manuscript). A two-dimensional Fourier transform was then taken to reconstruct the intensity distribution in the source plane. We reconstructed the picture using three different methods: (top right) based on the calculated CDC without noise as a benchmark, revealing the theoretical limits of the method; (bottom left) based on our Count scheme with photon number resolution in the detectors; (bottom right) based on the Traditional scheme. 

\setlength{\fboxsep}{0pt}
\setlength{\fboxrule}{2pt}

\begin{figure}[h!]
 \begin{center}
 \begin{tabular}{ccccc} 
 
 \fbox{\includegraphics[scale=0.45]{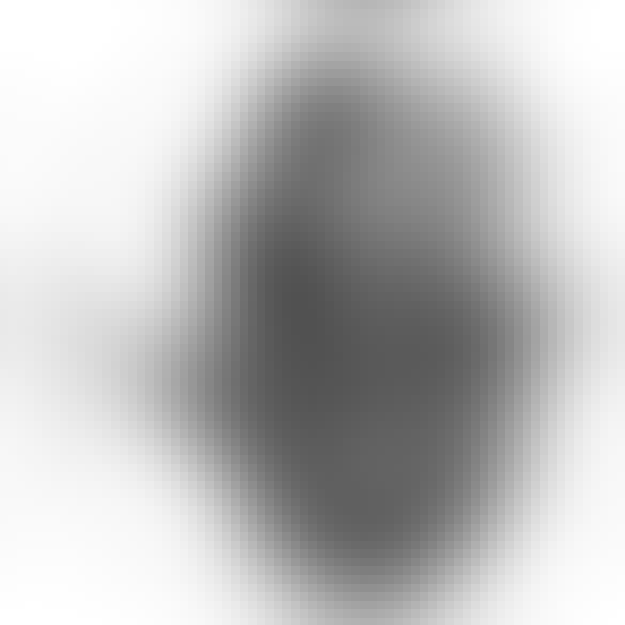}}  & 
 \hspace{-0.7cm} \vspace{0.1cm} \fbox{\includegraphics[scale=0.45]{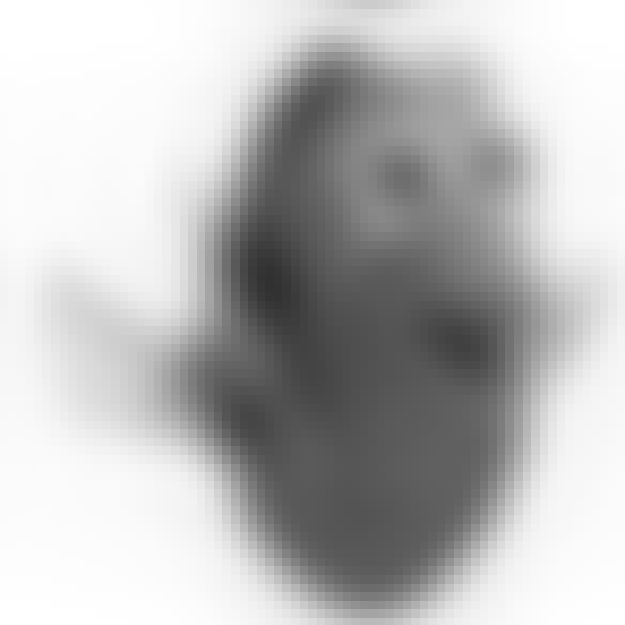}}    
 
  \fbox{\includegraphics[scale=0.45]{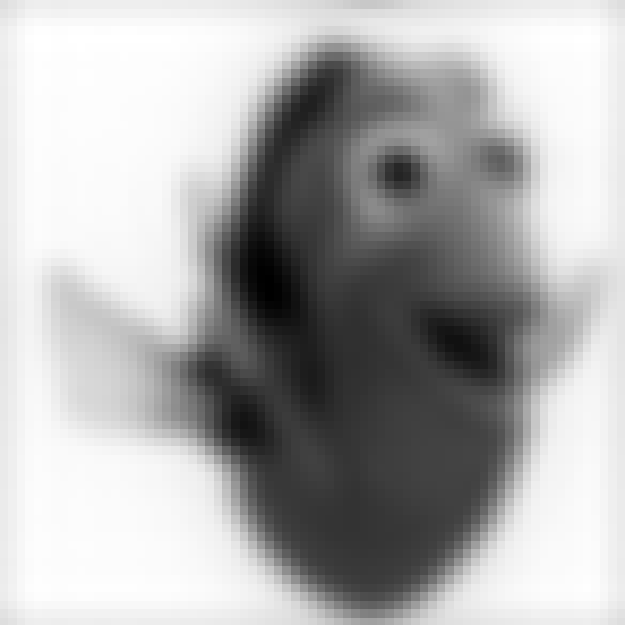}} & \hspace{-0.7cm} \fbox{\includegraphics[scale=0.45]{DoryD26.pdf}}  
  \fbox{\includegraphics[scale=0.45]{DoryOriginal.pdf}}\\ 
  
 \end{tabular}
 \caption{Simulated comparison of images reconstructed using arrays of detectors of increasing size. {Left}: image reconstruction using a 5$\times$5 detector array. {Second from left}: image reconstruction using a 10$\times$10 detector array.
 {Middle}: Image reconstruction using a 15$\times$15 detector array.
 {Second from right}: Image reconstruction using a 26$\times$26 detector array.
 {Right}: Original image before reconstruction section F. Imaging simulation.} 
 \label{imaging}
 \vspace{-9mm}
 \end{center}
 \end{figure}

\end{document}